\documentclass{article}
\usepackage[preprint]{spconfa4}

\usepackage[pdftex]{graphicx}
\usepackage{amsmath}
\usepackage{amssymb}
\usepackage{makecell}
\usepackage{array}
\usepackage{multicol}
\usepackage{multirow}
\usepackage{tikz}
\usepackage{anyfontsize}
\usepackage{lipsum}
\usepackage{marvosym}

\makeatletter
\newcommand\dlmu[2][4cm]{\hskip1pt\underline{\hb@xt@ #1{\hss#2\hss}}\hskip3pt}
\makeatother

\newcommand\blfootnote[1]{%
\begingroup
\renewcommand\thefootnote{}\footnote{#1}%
\addtocounter{footnote}{-1}%
\endgroup
}

\copyrightnotice{978-1-6654-3864-3/21/\$31.00 ©2021 IEEE}

\let\OLDthebibliography\thebibliography
\renewcommand\thebibliography[1]{
  \OLDthebibliography{#1}
  \setlength{\parskip}{0pt}
  \setlength{\itemsep}{0pt plus 0.3ex}
}

\begin{document}\sloppy

\title{Watermark Faker: Towards Forgery of Digital Image Watermarking}
\name{Ruowei Wang$^{\ast}$, Chenguo Lin$^{\ast}$, Qijun Zhao$^{\dagger}$, Feiyu Zhu$^{\dagger}$}
\address{
    College of Computer Science, Sichuan University, China\\
    \small\texttt{\{ruoweiwang,linchenguo\}@stu.scu.edu.cn, \{qjzhao,feiyuz\}@scu.edu.cn}
}

\maketitle \blfootnote{$^{\ast}$ Equal contribution. $^{\dagger}$ Co-corresponding authors.}

\begin{abstract}
Digital watermarking has been widely used to protect the copyright and integrity of multimedia data. Previous studies mainly focus on designing watermarking techniques that are robust to attacks of destroying the embedded watermarks. However, the emerging deep learning based image generation technology raises new open issues that whether it is possible to generate fake watermarked images for circumvention. In this paper, we make the first attempt to develop digital image watermark fakers by using generative adversarial learning. Suppose that a set of paired images of original and watermarked images generated by the targeted watermarker are available, we use them to train a watermark faker with U-Net as the backbone, whose input is an original image, and after a domain-specific preprocessing, it outputs a fake watermarked image. Our experiments show that the proposed watermark faker can effectively crack digital image watermarkers in both spatial and frequency domains, suggesting the risk of such forgery attacks.
\end{abstract}


\section{Introduction}\label{sec:intro}
\begin{figure}[!t]
	\centering
	\includegraphics[width=0.85\linewidth]{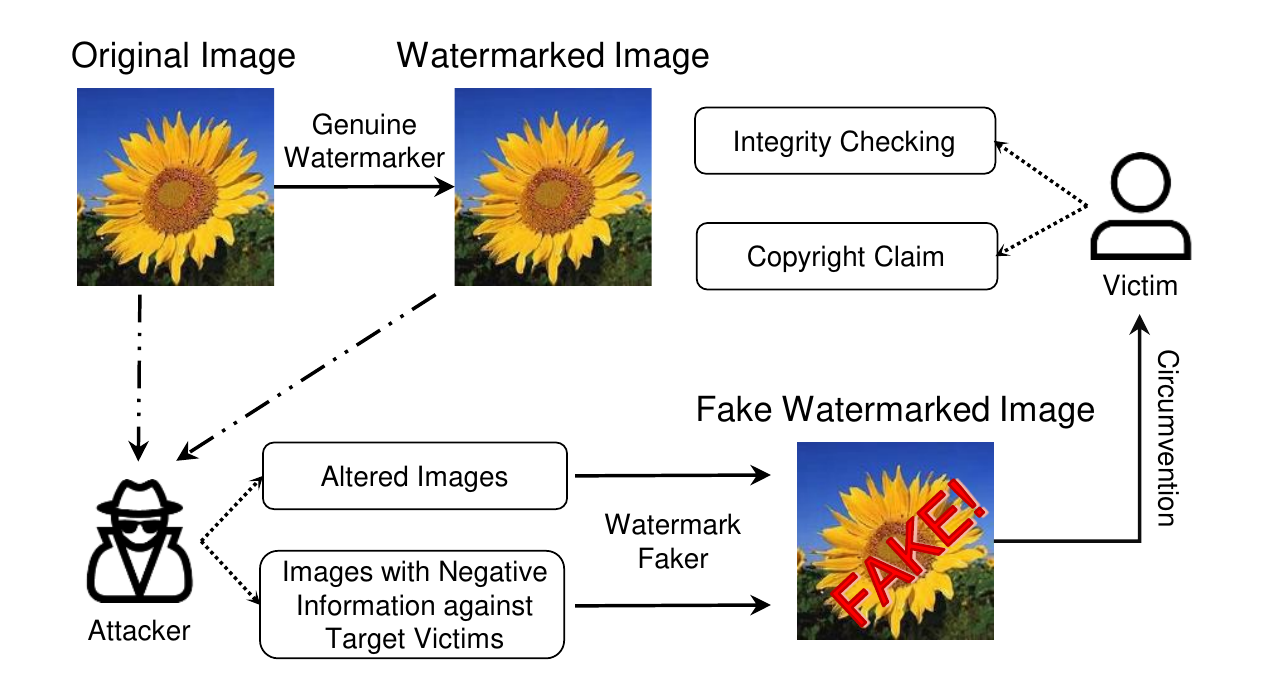}
	\caption{
	Illustration of application scenarios of forgery of digital image watermarking. The attacker could learn a watermark faker to circumvent target victims by using altered images or images with negative information against the victims.
	}
	\label{watermarkfaker}	
\end{figure}

The rapidly developed Internet and Web technology makes it easy to copy images and videos for unauthorised uses as well as to make fake information. As a technique for security, digital watermarking fights against those illegal acts in the fields of copyright protection and tamper detection. Recently, digital watermarking is combined with a lot of techniques to improve its ability in those fields, e.g. genetic algorithms~\cite{TRLG}, singular value decomposition~\cite{SVD_watermark}, deep learning~\cite{zhu2018hidden,zhang2020udh,deep_hiding_survey}, etc. In terms of the domain where watermarks hide, watermarking methods can be divided into spatial domain watermarking (e.g., LSB-based) and frequency domain watermarking (e.g., discrete cosine transform based)~\cite{block_dct_survey}. Previous studies on digital image watermarking mainly focus on improving the robustness against the attacks of destroying the embedded watermarks, but underestimate the risk of forgery attacks by generating fake watermarked images, with which attackers could infringe the copyright in watermarks or crack the tamper detection functionality (see Fig. \ref{watermarkfaker}).

The process of embedding watermarks into images is essentially a kind of image-to-image translation or image generation, which has been well developed in the past few years. Thanks to the emerging deep learning techniques, particularly generative adversarial networks (GAN)~\cite{pix2pix, pix2pixHD, SPADE, BicycleGAN, cycleGAN, starGAN, starGANv2, CoCosNet, Lee_2018_ECCV}. In~\cite{khan2020steganography}, Khan et al. made the first attempt to crack the least significant bit (LSB) based steganography~\cite{LSB} with GAN, though their method did not work well when fewer than five bits were used to embed secret messages (however, in practical usually if five or more bits are used, the changes to the original images will be visible). This motivates us to ask whether it is possible to generate fake watermarked images by using generative adversarial learning. 

\begin{figure*}[!t]
	\centering
	\includegraphics[width=\textwidth]{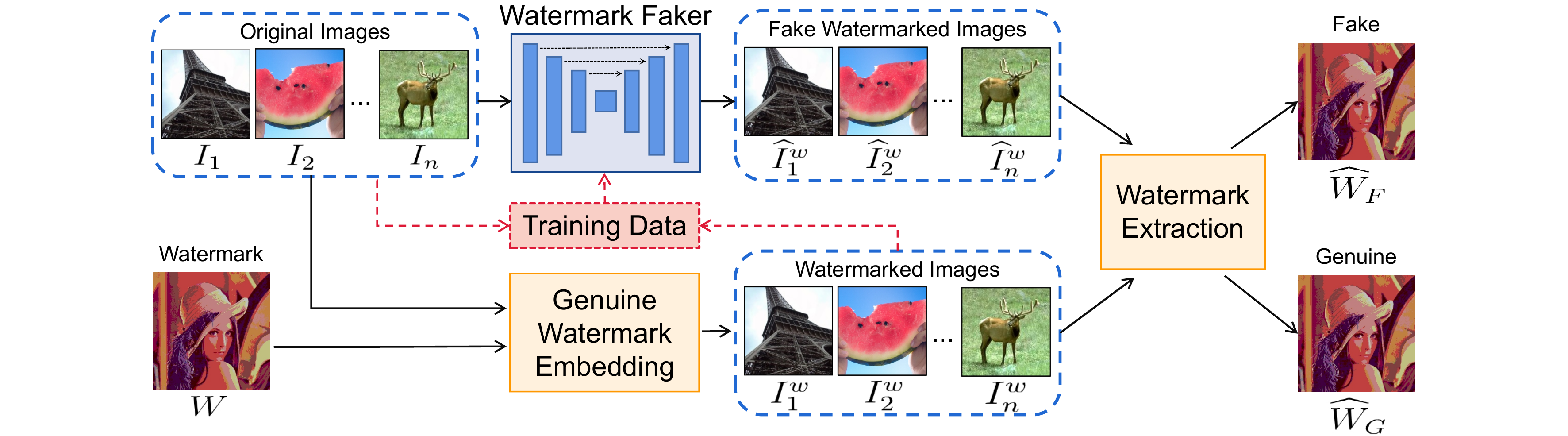}
	\caption{
	The basic idea of proposed forgery of a target digital image watermarking method.
	}
	\label{fig:basic_idea}	
\end{figure*}

This paper aims to make the first attempt to generate fake watermarked images to crack digital image watermarking. To this end, we construct a watermark faker by employing U-Net as the backbone and converting the original images to bit-wise representation as the input. Given a set of paired images of original and watermarked images generated by the watermarker under attack, we use these images to train the faker via adversarial learning. In the experiments, we consider three spatial domain watermarking methods and one frequency domain method, and the quantitative and qualitative results demonstrate the effectiveness of the proposed faker in forgery of digital image watermarking, which suggests the necessity of thoroughly studying such forgery attacks.

The rest of this paper is organized as follows. Section 2 introduces related work. Section 3 gives the detail of the proposed watermark faker, followed by evaluation experiments in Section 4. Finally, Section 5 concludes the paper.

\section{Related Work}\label{sec:related work}
\subsection{Digital Image Watermarking}
Digital image watermarking is an effective approach to address violations of information security in using multimedia data such as illegal copyright and tampering. It embeds into multimedia data with watermarks, which can be extracted or detected to make an assertion about the ownership or integrity of the data. Watermarking methods consist of two stages, embedding and extraction, i.e.,
\begin{equation}
    I^{w} = \mathcal{F}\left(I, W\right),\qquad
    \widehat{W} = \mathcal{G}\left(I^{w}\right),
    \label{embedding_extraction}
\end{equation}
where $I$ , $W$, $I^{w}$ and $\widehat{W}$ are, respectively, the original image, the watermark, the watermarked image and the extracted watermark. $\mathcal{F}$ and $\mathcal{G}$ denote the watermark embedding and extraction functions. In this paper, the watermark faker is a forgery of the embedding component of the watermark method under attack.

Watermarking methods can be divided into spatial domain watermarking and frequency domain watermarking~\cite{block_dct_survey}
Spatial domain watermarking methods embed watermarks directly by modifying some pixel values of original images. Frequency domain watermarking methods convert the original images to another domain (usually, frequency domain) before embedding. In this paper, we try to counterfeit three kinds of spatial domain watermarking methods, i.e., LSB~\cite{LSB}, LSB-M~\cite{LSB-M} and LSB-MR~\cite{LSB-MR}, and a frequency domain watermark, i.e., DCT-based watermarking~\cite{block_dct_survey}), which are widely used in protecting the copyright and integrity of multimedia data.

\subsection{Image-to-Image Translation}
Image-to-image translation is to learn a mapping from a source image domain to a target image domain~\cite{CoCosNet, Lee_2018_ECCV}.
Recently, many prominent image-to-image translation methods are inspired by conditional generative adversarial network (cGAN)~\cite{cGAN}. Considering the characteristics of used training data, we can split these methods into two categories. One takes paired data~\cite{pix2pix, pix2pixHD, SPADE, BicycleGAN}, while the other takes unpaired data~\cite{cycleGAN, starGAN, starGANv2, CoCosNet, Lee_2018_ECCV}. Generally, to learn a translation model from unpaired images is much more challenging. In this paper, we assume that paired data of original and watermarked images are available for training. Therefore, we design our watermark faker following the basic idea of pix2pix~\cite{pix2pix}, which is the first and highly influential framework for cGAN-based image-to-image translation.

\section{Method}\label{sec:method}

\begin{figure*}[!t]
	\centering
	\includegraphics[width=0.85\textwidth]{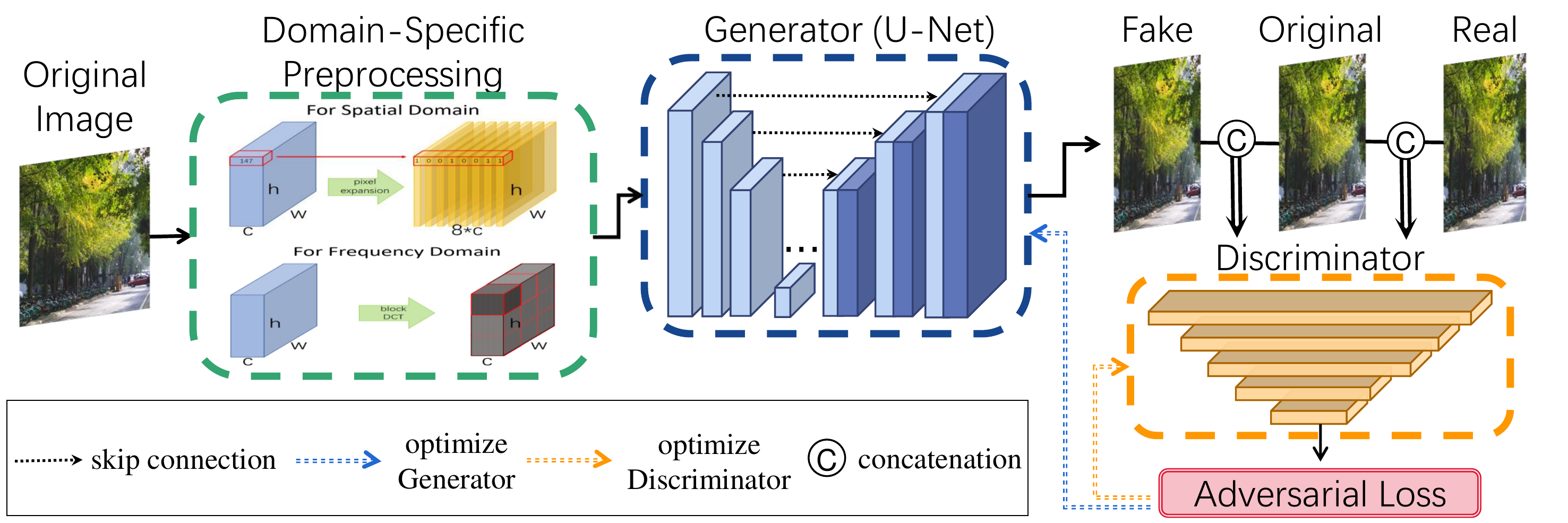}
	\caption{
	Diagram of Watermark Faker, the proposed forgery of digital image watermarking.
	}
	\label{fig:framework}	
\end{figure*}

\subsection{Overview of Proposed Watermark Faker}
Figure \ref{fig:basic_idea} depicts the basic idea of the proposed forgery of a target digital image watermarking method. As can be seen, a set of original images and the corresponding watermarked images generated by the to-be-attacked watermarking model are available for the attacker to train the watermark faker. Note that neither the embedded watermark nor the detail of the watermark embedding and extraction processes are known to the attacker. The trained watermark faker aims to generate fake watermarked images such that the watermarks extracted by the watermark extraction process from these fake images are close to the genuine watermark as much as possible.

The specific diagram of the implemented watermark faker is shown in Fig. \ref{fig:framework}. Given an original image, it is first undergoing some preprocessing such that the image is converted to proper representation in either spatial domain or frequency domain. The image is then translated into its watermarked version by a generator whose backbone is U-Net. This generator is trained via adversarial learning along with a discriminator based on PatchGAN~\cite{pix2pix}. Next, we introduce in detail the preprocessing and the adversarial learning employed by the proposed watermark faker.

\subsection{Domain-Specific Preprocessing}\label{sec:Domain-aware Preprocssing}

\textbf{Bit-wise representation: preprocessing for spatial domain watermarking.}\quad
The range of pixel values of digital images in spatial domain is determined by the number of bits used to represent the values. Because human vision systems are not sensitive to the least significant bits, spatial domain watermarking methods usually embed watermarks into these bits. As a result, if we directly compare the pixel values between real and fake watermarked images, the difference would be too minor to drive the training process (such issue is also known as gradient vanishing). This is essentially due to the fact that the impact of different bits varies according to their significance.

\begin{figure}[t]
	\centering
	\includegraphics[width=0.65\linewidth]{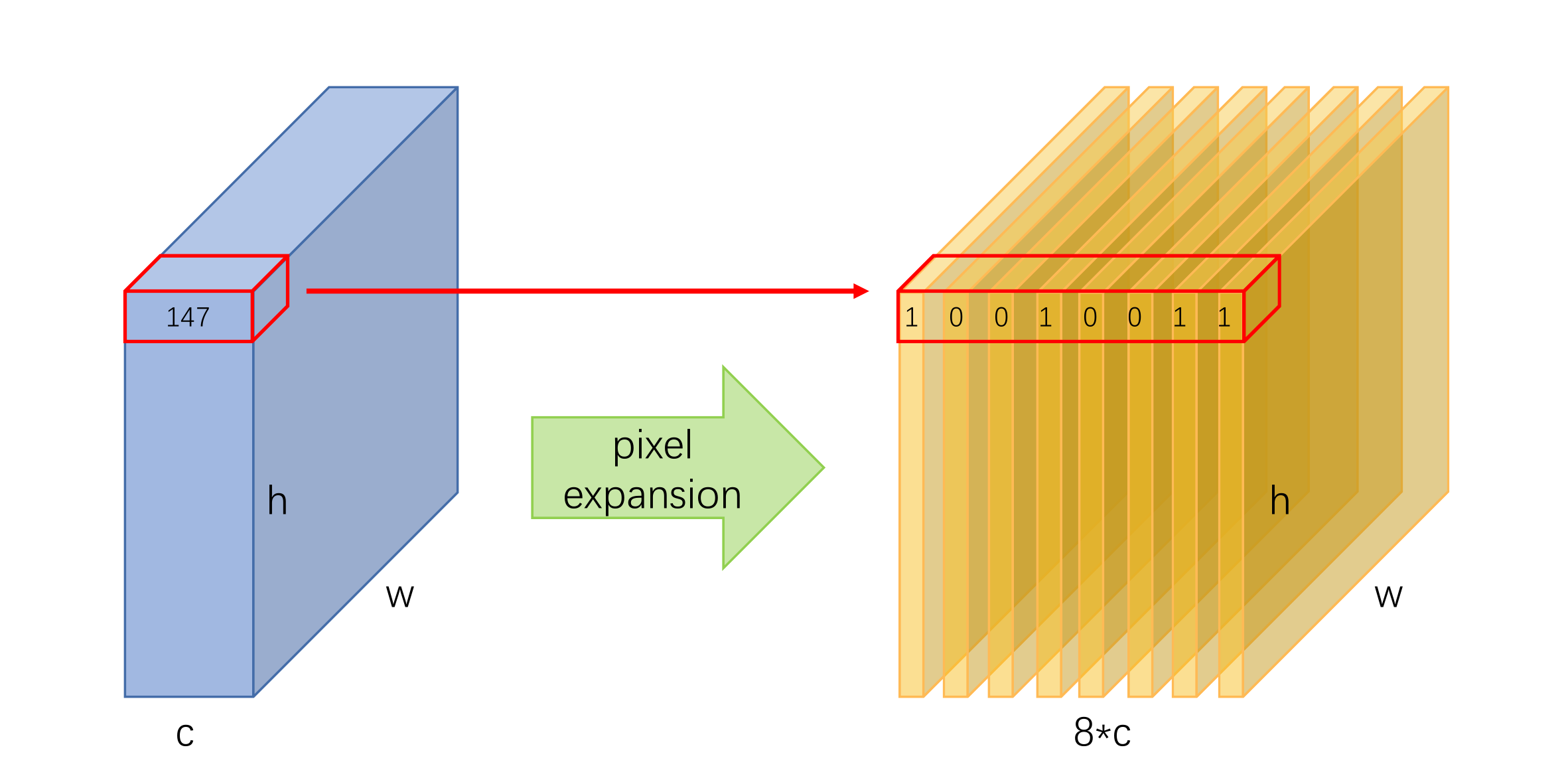}
	\caption{
	Pixel expansion operation converts an image to pixel-wise representations, in which each bit of pixel values corresponds to one channel
	}
	\label{Pixel Expansion}	
\end{figure}

To balance the contribution of different bits in the training process and better reveal the hidden pattern of watermark, we convert images to bit-wise representations by using Pixel Expansion (PE) as shown in Fig. \ref{Pixel Expansion}. For each pixel value $X$ of an image (say a eight-bit gray-scale image), we transform it from a decimal number to its corresponding binary form (eight bits):
\begin{equation}
    X = \Sigma_{i=0}^{L-1} 2^{i}\times x_{i},
    \label{equ:10to2}
\end{equation}
where $x_i$ denotes the $i^{th}$ bit of pixel value $X$, and $L$ is the number of bits used to represent pixel values. Then, $x_i$ of all $X$ in the image form a new channel. Finally, by translating a one-channel gray-scale image to an $L$-channel image, we get the bit-wise representation of the image. For a three-channel RGB image, we process each channel in the same way, resulting in a $(3\times L)$-channel image. By taking bit-wise representation of images as input, the faker can better learn the pattern of spacial domain watermarks.

\textbf{Preprocessing for frequency domain watermarking.}\quad
To achieve forgery of frequency domain watermarking, we implement the faker in frequency domain also. Specifically, we transform the original image into the frequency domain before applying the generator, and convert the result of the generator back to the spatial domain to obtain the fake watermarked image. In practice, since we have no prior of the specific transformation used by the watermarker under attack, we simply employ a blind search strategy among the typical image transformations between spatial and frequency domains.

\subsection{Adversarial Learning}
To train the watermark faker (more specifically, the generator $G$ in Fig. \ref{fig:framework}), we introduce a discriminator $D$ to implement adversarial learning between $G$ and $D$. As being motivated by the pix2pix image-to-image translation model~\cite{pix2pix}, we take the original image $I$ (in its preprocessed form) and a random noise vector $z$ (implemented in the form of dropout) as the input of $G$, and produce a synthetic image $G(I, z)$ as the fake watermarked image.

The discriminator $D$, working as a classifier, takes either a pair of synthetic image $G(I,z)$ and original image $I$ or a pair of target image $y$ and original image $I$ as input, and judges whether the image patches of $G(I,z)$ or $y$ are real. Here, we discriminate the image realness in local scale rather than in global scale, which enables the faker to better learn the detailed patterns of watermarks. Besides, we include the original image in the input, because the original image is an important reference to reveal the embedded watermark, especially when the watermark is related to the content of the original image (this is common in tamper-proof watermarking).

In adversarial learning, while $D$ is trying to distinguish the synthetic watermarked image $G(I,z)$ from the target real watermarked image $y$, $G$ is trained to do as well as possible in improving the quality of $G(I,z)$ to fool $D$. With the competition between $G$ and $D$, the generator learns a mapping from original images to the corresponding watermarked images. To fulfill the above learning process, we employ the following loss function, which is a conditional form of the least squares loss function inspired by CycleGAN~\cite{cycleGAN} and LSGAN~\cite{lsgan}.
\begin{equation}
\footnotesize
\label{equ:objective_function_type2_ls}
    \raggedleft
    \begin{aligned}
        \min _{D} \mathcal{L}(D) =&\mathbb{E}_{\boldsymbol{y}}\left[(D(\boldsymbol{I},\boldsymbol{y})-1)^{2}\right]+\mathbb{E}_{\boldsymbol{I},\boldsymbol{z}}\left[D(\boldsymbol{I},G(\boldsymbol{I},\boldsymbol{z}))^{2}\right],\\
        \min _{G} \mathcal{L}(G) =&\mathbb{E}_{\boldsymbol{I},\boldsymbol{z}}\left[(D(\boldsymbol{I},G(\boldsymbol{I},\boldsymbol{z}))-1)^{2}\right]+\mathbb{E}_{I,y,z}[||y-G(I,z)||_{1}].
    \end{aligned}
\end{equation}

\section{Experiments}\label{sec:exper}
\subsection{Data, Baselines and Metrics}
To evaluate the effectiveness of the proposed watermark faker, we randomly pick up $12,288$ images from Caltech256~\cite{griffin2007caltech} for training, and $2,048$ images from Caltech256 for testing. Three LSB-based watermarking methods (LSB~\cite{LSB}, LSB-M~\cite{LSB-M}, and LSB-MR~\cite{LSB-MR}) and one DCT-based watermarking method ($8\times8$ block DCT -based watermarking~\cite{block_dct_survey}) are chosen as target watermarkers to attack. Since this work is the first for forgery of digital image watermarking, we cannot find any existing watermark fakers for direct comparison. Hence, we choose pix2pix, which is a widely-used image-to-image translation model and contributes part of the backbone of our network, as the baseline. For quantitative comparison, we take PSNR\cite{ssim-psnr} and SSIM~\cite{ssim} as metrics to compute the similarity between the fake watermarked images as well as the extracted fake watermarks (i.e., $\widehat{I}^{w}$ and $\widehat{W}_{F}$) and the corresponding real ones (i.e., ${I}^{w}$ and $\widehat{W}_{G}$). The higher the metric values are, the more effective the watermark fakers are.

\begin{figure}[!t]
    \centering
    \includegraphics[width=0.9\linewidth]{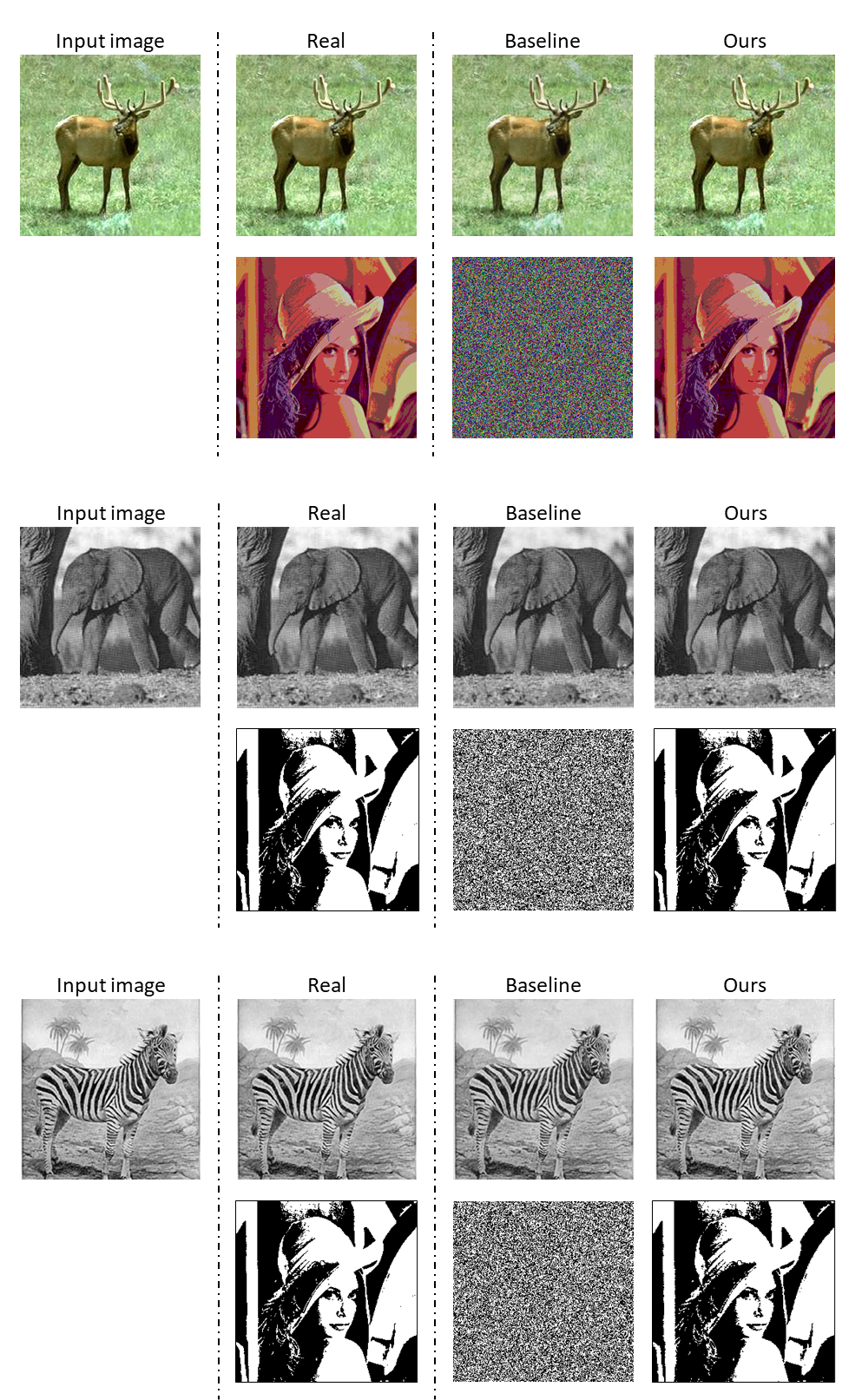}
    \caption{Some forgery results of different methods for three spatial domain watermarking methods. In each sub-figure, the first column shows the input original image, the second row shows the real watermarked image and the extracted real watermark, and the third and forth columns show the forgery results of the baseline and our proposed methods, respectively.
}
\label{fig:experiments}
\end{figure}

\subsection{Forgery Results of Spatial Domain Watermarking}
Fig. \ref{fig:experiments} shows some fake watermarked images generated by the baseline pix2pix and the proposed methods in contrast to the corresponding real watermarked images. As can be seen, visually, the watermarked images generated by the two methods both appear quite similar to the real ones; however, in terms of the extracted watermarks, the baseline pix2pix method results in pure noise, whereas the proposed method successfully counterfeits the watermark. 

Quantitative results in terms of PSNR and SSIM are presented in Table \ref{table:lsb}. Although the baseline method achieves a little bit higher similarity between fake and real watermarked images when attacking LSB, the proposed method performs substantially better in generating fake watermarks when attacking all the three watermarkers. These results prove that the proposed method is much more effective in forgery of spatial domain image watermarks while keeping the fake watermarked images visually plausible.

\begin{table*}[!t]
\renewcommand{\arraystretch}{1}
\caption{The average PSNR and SSIM values of different forgery methods for spatial domain watermarking.
}
\label{table:lsb}
\footnotesize
\centering
\begin{tabular}{c|cccc|cccc|cccc}
    \hline
    \multirow{3}{*}{} 
    & \multicolumn{4}{c|}{\dlmu[2.7cm]{\textcolor{white}{g}LSB\textcolor{white}{g}}}
    & \multicolumn{4}{c|}{\dlmu[2.7cm]{\textcolor{white}{g}LSB-M\textcolor{white}{g}}}
    & \multicolumn{4}{c}{\dlmu[2.7cm]{\textcolor{white}{g}LSB-MR\textcolor{white}{g}}}
    \\
    
    & \multicolumn{2}{c}{\dlmu[1.6cm]{Image}}
    & \multicolumn{2}{c|}{\dlmu[1.6cm]{\textcolor{white}{g}Watermark\textcolor{white}{g}}}
    & \multicolumn{2}{c}{\dlmu[1.6cm]{Image}} 
    & \multicolumn{2}{c|}{\dlmu[1.6cm]{\textcolor{white}{g}Watermark\textcolor{white}{g}}}
    & \multicolumn{2}{c}{\dlmu[1.6cm]{Image}} 
    & \multicolumn{2}{c}{\dlmu[1.6cm]{\textcolor{white}{g}Watermark\textcolor{white}{g}}} 
    \\
    & PSNR & SSIM 
    & PSNR & SSIM 
    & PSNR & SSIM 
    & PSNR & SSIM 
    & PSNR & SSIM 
    & PSNR & SSIM
    \\ 
    \hline
    Baseline         
    & \textbf{34.204}
    & \textbf{0.953}
    & 8.789
    & 0.009
    & 35.096
    & 0.965
    & 3.013
    & 0.001
    & 33.246
    & 0.965
    & 3.010
    & 0.001
    \\
    Ours        
    & 30.963
    & 0.941
    & \textbf{36.805}	
    & \textbf{0.986}	
    & \textbf{44.555}	
    & \textbf{0.987}	
    & \textbf{42.112}	
    & \textbf{0.999}	
    & \textbf{40.659}
    & \textbf{0.981}	
    & \textbf{20.702}
    & \textbf{0.953}
    \\
    \hline
\end{tabular}
\end{table*}

\begin{table*}[!t]
\renewcommand{\arraystretch}{1}
\caption{The average PSNR and SSIM values of different forgery methods for frequency domain watermarking.
}
\label{ablation:dct}
\footnotesize
\centering
\begin{tabular}{c|cc|cccc}
	\hline
	\multicolumn{1}{c|}{\multirow{3}{*}{}} 
	& \multicolumn{2}{c|}{\dlmu[2cm]{\textcolor{white}{g}Image\textcolor{white}{g}}}            
	& \multicolumn{4}{c}{\dlmu[3.5cm]{\textcolor{white}{g}Watermark\textcolor{white}{g}}}                    
	\\
	\multicolumn{1}{c|}{}                  
	& \multirow{2}{*}{PSNR} 
	& \multirow{2}{*}{SSIM} 
	& \multicolumn{2}{c}{\dlmu[2.7cm]{\textcolor{white}{g}Real$\longleftrightarrow$ GT\textcolor{white}{g}}}
	& \multicolumn{2}{c}{\dlmu[2.7cm]{\textcolor{white}{g}Fake$\longleftrightarrow$ GT\textcolor{white}{g}}}
	\\
	\multicolumn{1}{c|}{}                  
	&
	& 
	& PSNR
	& \multicolumn{1}{c}{SSIM} 
	& PSNR
	& SSIM
	\\ 
	\hline
	\multicolumn{1}{c|}{Baseline}              
	&26.809029                   
	&0.848049           
	&17.271548
	& \multicolumn{1}{c}{0.750439} 
	&4.693388           
	&0.117471           
	\\
	\multicolumn{1}{c|}{Ours}              
	& \textbf{32.668204}                  
	& \textbf{0.920432}                 
	& 17.271548
	& \multicolumn{1}{c}{0.750439} 
	& \textbf{9.772256}	       
	& \textbf{0.545196}	 
	\\
	\hline
\end{tabular}
\end{table*}

\begin{figure}[!t]
    \centering
    \includegraphics[width=0.75\linewidth]{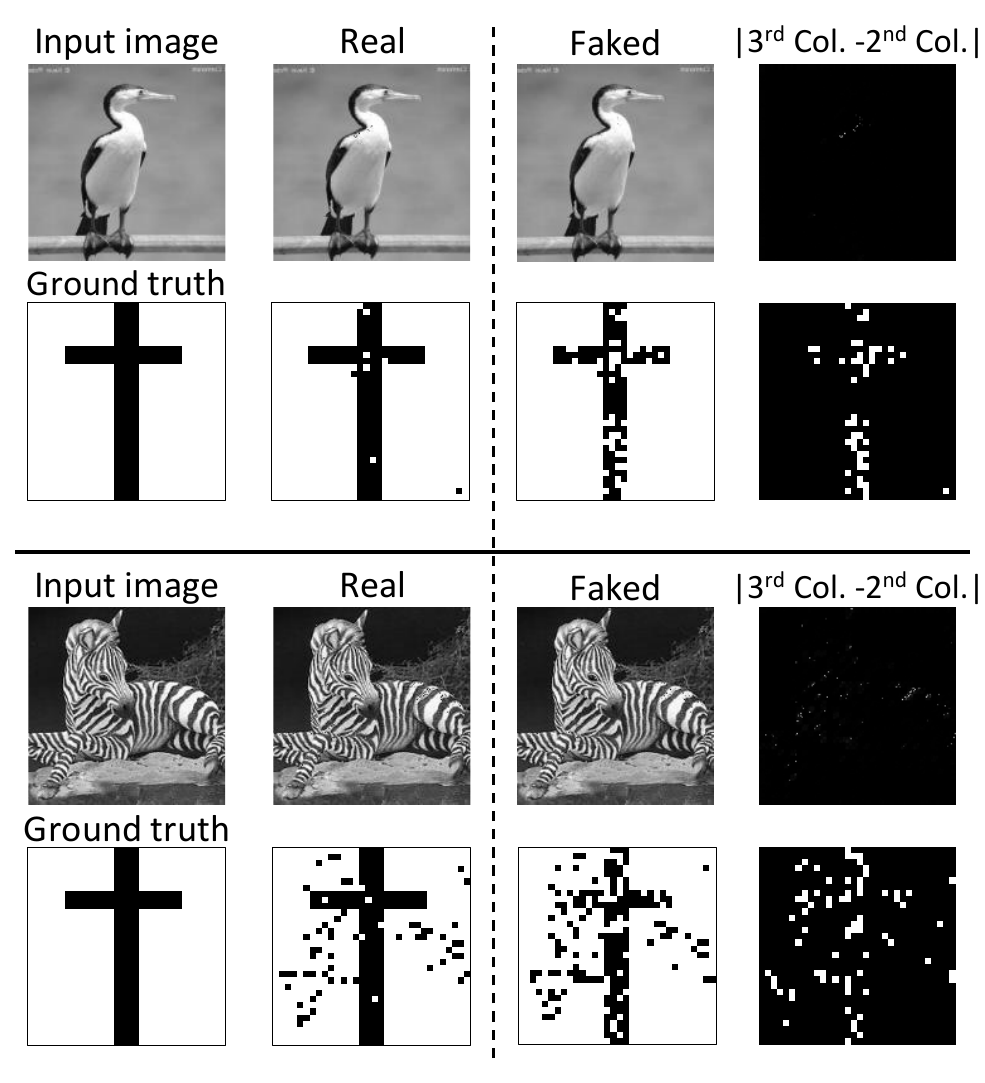}
    \caption{
    Example forgery results of the proposed method for DCT-based frequency domain watermarking.
    }
    \label{fig:the quality of image, DCT}
\end{figure}

\subsection{Forgery Results of Frequency Domain Watermarking}
For frequency domain watermarking, because the extracted real watermarks are usually not exactly same as the watermarks embedded into the original images (see Fig. \ref{fig:the quality of image, DCT}), we compute the PSNR and SSIM metrics between the extracted watermarks and the embedded ones, and then compare the metric values obtained by the proposed watermark faker with those obtained by the to-be-attacked target watermarker. The goal of the watermark faker is to make its metric values as close to the metric values of the target watermarker as possible. Table \ref{ablation:dct} gives average PSNR and SSIM values of the proposed watermark faker and the target watermarker. From these results, we can see that frequency domain watermarks are more difficult to counterfeit than spatial domain watermarks; nevertheless, as shown in Fig. \ref{fig:the quality of image, DCT}, the proposed method can still to some extent learn the pattern of embedded watermarks. 

\subsection{Ablation Study}
In this section, we compare some different implementations of the proposed watermark faker to assess (i) how much the pixel expansion (PE) operation contributes to the forgery of spatial domain watermarking, and (ii) what if directly generating fake watermarked images in the spatial domain when attacking frequency domain watermarking. The results are summarized in Tables \ref{ablation:lsb} and \ref{ablation:dct}, from which we can observe that (i) for the spatial domain watermarker (i.e., LSB), applying PE during preprocessing can significantly improve the effectiveness of counterfeiting watermarks, though the visual quality of the generated fake watermarked images becomes somehow worse (but still acceptable); (ii) for the frequency domain watermarker (i.e., DCT-based), the similarity between extracted fake watermark and the embedded watermark (taken as the ground truth) is obviously closer to that between extracted real watermark and the ground truth, which suggesting that it is more effective to crack DCT-based watermarking in frequency domain than in spatial domain.

\begin{table}[!t]
\renewcommand{\arraystretch}{1}
\caption{Ablation study results of forgery of spatial domain watermarking.}
\label{ablation:lsb}
\footnotesize
\centering
\begin{tabular}{c|cccc}
	\hline
	\multicolumn{1}{c|}{}            
	& \multicolumn{4}{c}{\dlmu[3cm]{\textcolor{white}{g}LSB\textcolor{white}{g}}}
	\\
	& \multicolumn{2}{c}{\dlmu[1.8cm]{\textcolor{white}{g}Image\textcolor{white}{g}}}
	& \multicolumn{2}{c}{\dlmu[1.8cm]{\textcolor{white}{g}Watermark\textcolor{white}{g}}}
	\\
    & PSNR
	& SSIM
	& PSNR
	& SSIM
	\\ 
	\hline
    Ours without PE
	&\multicolumn{1}{c}{\textbf{34.204748}}
	& \multicolumn{1}{c}{\textbf{0.953002}}
	&8.789933
	&0.009935       
	\\
    Ours with PE  
	&\multicolumn{1}{c}{29.209504}
	& \multicolumn{1}{c}{0.911588} 
	&\textbf{27.118629}
	&\textbf{0.880872}         
	\\        
	\hline     
\end{tabular}
\end{table}

\section{Conclusions}\label{sec:concl}
This paper for the first time shed light on the new open issue of forgery of digital image watermarking using latest deep learning based image generation technology. With U-Net as the backbone, we construct a watermark faker, and via adversarial learning, we get the faker trained with respect to a specific target watermarker. Our quantitative and qualitative evaluation results demonstrate that the proposed watermark faker can effectively counterfeit digital watermarks in both spatial and frequency domains. Although existing forgery results we obtain for frequency domain watermarking are not yet as good as the results for spatial domain watermarking, our study in this paper reveals the potential risk caused by forgery of digital watermarking with deep learning technology, which is however seriously under-estimated by contemporary researchers and practitioner in the field of digital watermarking. In the future, we are going to investigate the possibility of cracking more advanced watermarking methods and to study how to detect forgery of digital watermarks.

\bibliographystyle{IEEEbib}
\bibliography{mybib}

\end{document}